\documentclass[%utf8%,draft % раскомментировать, если нужен черновой режим
               ]{jetp4} % jetp.cls
\twocolumn % закомментировать, если не нужен двухколоночный вариант
\addto\captionsenglish{\renewcommand{\refname}{REFERENCES}}
\let\figori=\fig \def\fig[#1]#2{\figori[#1]{*}}

\begin{document}

\title{Comment on the article  ''Non-local gravitational corrections in black hole
shadow images'' by S. O. Alexeyev et al. }

\rtitle{Comment on the article  ''Non-local gravitational corrections \dots''}

\author{Alexander F.}{Zakharov} % (a,б)

\affiliation{National Research Center  -- ''Kurchatov Institute'', Moscow} % a
\affiliation{Bogoliubov Laboratory of Theoretical Physics, JINR, 141980 Dubna, Russia
             } % б

\abstract{
Recently Alexeyev et al. published paper (J. Theor. Exper. Phys. v. 165, N 4, p. 508  in Russian;  	arXiv:2404.16079 [gr-qc], the reference is given also in \cite{Alexeyev_24}).
   In the paper the authors discussed an opportunity of estimating spins from the analysis of the shadow reconstruction of black holes,
theoretically considered using the nonlocal gravity model proposed earlier for the description of ''quantum'' black holes. However, in essence, this paper
considered circular photon orbits, and the fact that the corresponding motion parameters determine the shape and size of shadows, similarly to Kerr black holes, remained unproven.
It is also remained unproven the statement that for an equatorial observer the shadow size in the direction of rotation of ''quantum'' black holes remains independent of spin. A long time ago the shadow property was established for the Kerr black hole case.
}

\maketitle

Many years ago it was shown{\cite{Zakharov_86} that if we consider the constants of motion for classical Kerr black holes, the capture region and the scattering region for photons are separated by the Chandrasekar constants of motion ($\xi, \eta$) corresponding to circular photon orbits. Thus, for the Kerr metric, the shape and size of the shadow
is determined by these critical parameter values (as shown in \cite{Zakharov_05}).
As is known, the possibility of shadow reconstruction in the neighborhood of the nearest supermassive black holes is currently under discussion,
not only for classical Kerr--Newman black holes, but also for some of their ''quantum'' generalizations, although in some cases quantum corrections are used to recover shadows in the vicinity of the nearest supermassive black holes.
in some cases quantum corrections in the corresponding coefficients are too small for their influence on the physical effects to be detected (this is also noted by the authors of the paper \cite{Alexeyev_24}).

If we mean a purely theoretical discussion, we can analyze the differences of shadows for the classical Kerr black hole and
its ''quantum'' generalization, considered in the work of \cite{Alexeyev_24}, but it is necessary to keep in mind that if we speak about astrophysical black holes, it is necessary to take into account the influence of such factors as the spatial mass distribution, the influence of plasma effects, etc., since the influence of these factors
significantly exceeds the difference in the shape and size of shadows for the cases of a classical black hole and its quantum generalization.
In paper \cite{Zakharov_05} it is shown that for a classical Kerr black hole in the case of the observer's position in the equatorial plane,
the size of the shadow in the direction of the black hole's rotation does not depend on the spin of the black hole. The authors of the paper \cite{Alexeyev_24} note that in the examples considered by them
the size of shadows for the ''quantum'' generalization of the Kerr black hole for an observer in the equatorial plane is also independent of spin, however
it remains unproven that for additional parameters (due to the use of the model of nonlocal gravitation),
the sizes of shadows in the rotation direction for the considered ''quantum'' generalization of the Kerr black hole for an observer in the equatorial plane, do not depend on spin.

After the discovery of any physical (or astronomical) phenomenon, there is also its theoretical explanation, but very often theoretical predictions are not realized in experiments or astronomical observations.
are realized in experiments or astronomical observations, so it is useful to recall that the  
the idea to use ground-based and ground-based space-based VLBI operating in the millimeter or submillimeter range to reconstruct the shadow in the vicinity of the Galactic center was proposed in \cite{Zakharov_05} (which can naturally be generalized to other supermassive black holes, such as the black hole at the center of the galaxy M87). 
The possibility of reconstructing the shadow of a black hole at the Galactic Center using global ground-space (and ground-based) interferometers operating in the mm band was {\it firstly} proposed in  \cite{Zakharov_05} that was not mentioned in  \cite{Alexeyev_24}. 
  This prediction of paper \cite{Zakharov_05} paper was remarkably fully confirmed when the Event Horizon Telescope (EHT) collaboration reconstructed the size of the shadow at the Galactic Center (GC) using the ground-based global interferometer  VLBI  operating in the millimeter \cite{Akiyama_22a}. 
 Thus, J. Bardeen's thought experiment (mentioned below)
  essentially became another test of the general theory of relativity \cite{Zakharov_24}.

 This fact of prediction of the possibility of shadow detection in the Galactic Center is known, so, in particular, in a recent paper \cite{Bambhaniya_24} it is noted that the idea of recovering the shadow of a black hole in the center of the Galaxy with the help of global interferometers working in the millimeter wavelength range was originally suggested in the paper \cite{Zakharov_05}.
As noted earlier, the prediction of what object should be observed, by what means, and what should result from the observations is not often realized exactly, as was the case with the prediction of the possibility of shadow recovery in the GC realized by the EHT collaboration.
Thus, it is important to note the fact that the prediction about the possibility of black hole shadow recovery at the Galactic Center was realized and 
an estimate of the shadow size for Sgr A*, which, as predicted in \cite{Zakharov_05} and confirmed by the EHT collaboration \cite{Akiyama_22a}, is about 50 angular microseconds.

About half a century ago, J. M. Bardeen considered a thought experiment in which there is a flat radiating screen behind a black hole with extreme rotation ($\hat{a}=1$) at infinity
\cite{Bardeen_73} and then a distant observer located in the equatorial plane can observe a dark spot, and it is assumed that
photons move along geodesics and are not scattered (e.g. on electrons) in the neighborhood of the black hole.
A similar pattern was subsequently reproduced in Chandrasekar's book \cite{Chandra_83}.
However, neither Bardeen nor Chandrasekar discussed the possibility of observational detection of such a dark spot (shadow) because
a) even for supermassive black holes, the sizes of shadows are very small;
b) as a rule, there is no luminous screen behind an astrophysical black hole;
c) it is rather difficult to distinguish the dark region (shadow) from the area with low luminosity. 
Answers to the questions of how one can reconstruct the shadow from observations were answered in \cite{Zakharov_05}. Indeed, in the 2000s, space-ground interferometers such as Radioastron (whose angular resolution at the shortest wavelength of 1.3 cm was on the order of 7 angular microseconds) and Millimetron (with angular resolution several orders of magnitude better) had already been discussed, and the ground-based global VLBI systems had angular resolutions of a few tens of angular microseconds,
 which is comparable to the size of the shadow at the Galactic Center and M87*. In the presence of radiation sources outside the photon circular orbits, the secondary images of these sources should be close to the shadow and this makes it possible to recover the shadow of the black hole, as was predicted in \cite{Zakharov_05} and realized in \cite{Akiyama_22a} when recovering the shadow of the black hole located in the Galactic center.

Let us recall the definition of photon circular orbits.
Let's consider the case of a photon moving from infinity to a Kerr black hole. In this case, the motion of the photon is determined
polynomial $\hat{R}_{ph}(\hat{r})$ \cite{Zakharov_91,Zakharov_94}.
 As shown in the work \cite{Zakharov_86}, if the polynomial $\hat{R}_{ph}(\hat{r})$ has no roots for the value
$\hat{r} > \hat{r}_+=1+\sqrt{1-\hat{a}^2}$, then the photon is captured by the black hole if the polynomial $\hat{R}_{ph} (\hat{r})$ has a single root
$ \hat{r}_t > \hat{r}_+$ $(\hat{R}_{ph}(\hat{r}_t)=0)$ (and $\left.\dfrac{\partial {\hat{R}_{ph}}}{\partial {\hat{r}}}\right|_{\hat{r}=\hat{r}_t} > 0$),
then there is scattering of a photon on a black hole and the approach of the photon to the black hole is replaced by removal and
in case there is a double root of the polynomial $\hat{R}_{ph}(\hat{r_s})=\left.\dfrac{\partial{\hat{R}_{ph}}}{\partial { \hat{r}}}\right|_{\hat{r}=\hat{r}_s}=0$, then a photon moving from infinity approaches an orbit with a constant value of the radial coordinate
$\hat{r}_s=const$
and these orbits in the literature are often called {\it circular}, although they are exactly circular only in the case of equatorial motion at $\eta=0$.

In case of consideration of generalizations of the classical Kerr black hole metric, the authors of the paper \cite{Alexeyev_24} would have to prove that for the region of values of the aiming parameters
inside the region, which they call ''shadow'', there really occurs the capture of photons by a ''quantum'' black hole obtained using the nonlocal theory of gravitation, i.e. it was necessary to generalize the statements of the paper \cite{Zakharov_86} to the case of the black hole metric with ''quantum'' corrections considered by the authors of \cite{Alexeyev_24}. In the absence of proof of the above statement
for the black hole metric with ''quantum'' corrections considered in paper \cite{Alexeyev_24} we can speak only about the impact parameters corresponding to circular photon orbits,
and that these regions of constants of motion separate the regions of capture and scattering (as it happens for Kerr black holes)
 remains an unproved assumption and thus it remains unproved that the motion parameters corresponding to circular photon orbits determine the shadow boundary.
 In the paper \cite{Zakharov_05} it is noted that the curve $\eta(\xi)$, defining the set of values of the constants of motion $(\eta,\xi)$, corresponding to the twofold root of the polynomial
 $\hat{R}_{ph}(\hat{r})$ and the maximum value of the function $\eta(\xi)$ equal to 27 is taken at the value $\xi=2a$, i.e., $\eta(2a)=27$. In this case, at the equatorial position of the distant observer the shadow boundary is determined by the curve $\beta(\alpha)$ (in accordance with the notations of the paper \cite{Zakharov_05}), and the maximum value of the function $\beta(\alpha)$, equal to $3\sqrt{3}$ is taken at the value $\alpha=2a$, thus, the size of the shadow in the direction of rotation of the black hole in the case of the equatorial position of the distant observer does not depend on the value of the spin $a$.
 of spin $a$. This fact is used by the authors of the \cite{Alexeyev_24} both for the Kerr metric and for the case of the black hole metric with the 
 ''quantum'' corrections.
As noted above, in \cite{Zakharov_05} it is proved that in the case of a Kerr black hole and equatorial position of a distant observer, the size of the black hole's shadow in the spin direction does not depend on spin and is equal to $2\sqrt{27}M$. This fact is used by the authors of the \cite{Alexeyev_24} as a known statement, but the corresponding reference is not given. 
the corresponding reference is not given. 

 Thus, in the paper \cite{Alexeyev_24} it is used the fact that for an equatorial observer the shadow size in the spin direction of the considered by them ''quantum generalization'' of the Kerr black hole (''Examples 1 -- 4'') does not depend on spin, that is only illustrated by the given examples with some chosen values of parameters $\alpha$ and $\beta$, but it remains unknown for what region of values of the parameters $\alpha$ and $\beta$ the statement about independence of the shadow size in the direction of rotation of the black hole remains valid for the case of a ''quantum'' black hole obtained using the model of nonlocal gravitation.

If we compare Figure 2 from the \cite{Zakharov_05} (where the spin values $a=0, a=0.5, a=1$ are considered) and Figure 2 (top panel) from the \cite{Alexeyev_24}
(where the values of spin $a=0, a=0.3, a=0.5, a=0.9, a=0.98$ are considered), one can come to the conclusion that these figures coincide to a large extent, thus discussing essentially the same properties of shadows ($a=0, a=0.3, a=0.5, a=0.9, a=0.98$).
the same properties of the shadows (change in the shape of the shadows and the size of the shadow in the spin direction, independent of the spin magnitude), i.e., in the later paper \cite{Alexeyev_24}
it is not noted that in the above case the properties of shadows, known and discussed in the literature already for 20 years, are discussed.

 In paper \cite{Alexeyev_24}
the authors  discussed the shadow of a ''quantum'' black hole obtained using the nonlocal gravity model, the differences in the shape and size of the shadow with increasing values of the ''quantum corrections'' by many orders of magnitude were noted. In this connection the question arises. What physical (astronomical) formulation of the problem in the paper \cite{Alexeyev_24} reflects the black hole metric with such increased ''quantum corrections'' (because in this case the ''quantum'' corrections are increased by many orders of magnitude and at the same time such natural astronomical corrections  are ignored, for example, the distribution of matter near the black hole is neglected, while it could change the gravitational field and thus the shape and size of the shadow)?

%\small

{}


\begin{thebibliography}{}
%\small
\bibitem{Alexeyev_24}
S. O. Alexeyev, A. A. Baiderin, A. V. Nemtinova, O.I. Zenin, 
%Нелокальные гравитационные теории и
%изображения теней черных дыр, 
J. Theor. Exp. Phys. {\bf 165},  508 (2024) (in Russian); arXiv:2404.16079 [gr-qc].


\bibitem{Zakharov_86}
A. F. Zakharov,  %О типах неограниченных орбит в метрике Керра
Sov. Phys. JETP, {\bf 64}, 1 (1986).


\bibitem{Zakharov_05} 
{A. F. Zakharov, A. A. Nucita, F. De Paolis, G.  Ingrosso}, % {Measuring the black hole parameters in the galactic center
%with RADIOASTRON}~//
%%\href{https://doi.org/10.1016/j.newast.2005.02.007}
 {New Astron.} {\bf 10},  479 (2005);
arXiv:astro-ph/0411511.

\bibitem{Zakharov_24}
A. F. Zakharov, %“Shadows near supermassive black holes: From a theoretical concept to GR test,” 
Intern. J. Mod. Phys. D {\bf 54}, 2340004 (2023). 
%\url{https://doi.org/10.1142/S0218271823400047}.
arXiv:
2308.01301.

\bibitem{Akiyama_22a}%
      {K. Akiyama,  A.  Alberdi, W.  Alef
  %\emph
   et~al.} (Event Horizon
  Telescope Collaboration),
 % {First Sagittarius A* Event Horizon Telescope Results.
%  I. The Shadow of the Supermassive Black Hole in the Center of the Milky Way}~//
%     %\href{https://doi.org/10.3847/2041-8213/ac6674}
     {Astrophys. J. Lett.}, {\bf 930}, L12 (2022).
  
\bibitem{Bambhaniya_24}
P. Bambhaniya, A. B. Joshi, D. Dey et al.
%Relativistic orbits of S2 star in the presence of scalar field,
Eur. Phys. J. C {\bf 84}, 124 (2024).

%\bibitem{Carter_68}
%{B. Carter}, % {Global Structure of the Kerr Family of Gravitational Fields}~//
%%%\href{https://doi.org/10.1016/j.newast.2005.02.007}
%{Phys. Rev.} {\bf 174},  1559 (1968).





\bibitem{Zakharov_91}
{A. F. Zakharov}, %Орбиты фотонов и  ультрарелятивистских частиц в гравитационном поле вращающихся черных дыр
Sov. Astron. {\bf 35},  30 (1991).

\bibitem{Zakharov_94}
A. F. Zakharov, %On the hotspot near a Kerr Black hole: Monte Carlo simulations
Mon. Not. R. Astron. Soc. {\bf 269}, 283 (1994).

%\bibitem{Zakharov_99}
%{A. Ф. Захаров, С. В. Репин}, %Спектр аккреционного диска в окрестности вращающейся черной дыры
%Астрон. ж., {\bf 76}, 803 (1999).

%\bibitem{Mino_03}
%Y. Mino, %Perturbative approach to an orbital evolution around a supermassive black hole
%Phys. Rev. D, {\bf 67}, 084027 (2003).
%
%
%\bibitem{Hackmann_19}
%E. Hackmann, Probing General Relativity
%in the strong gravity regime, Habilitation Thesis, Dem Fachbereich Physik und Elektrotechnik der Universit\"at
%Bremen zur Habilitation f\"ur das Lehr- und Forschungsgebiet, Theoretische Physik vorgelegte wissenschaftliche Arbeit,
%Bremen University, p. 23 (2019), \url{https://www.evahackmann.org/research/}.

\bibitem{Bardeen_73}
 J. M. Bardeen, in {\it Black Holes
    (Les Astres Occlus)},     eds. 	B. S. DeWitt and C. DeWitt-Morette (New York: Gordon \& Breach, 1973)  p. 215.

\bibitem{Chandra_83}
S. Chandrasekhar, {\it Mathematical Theory of Black Holes},
Clarendon Press, Oxford (1983).


%\bibitem{Lemaitre_31}
%G. Lemaitre,  The expansing Universe, Mon. Not. R. Astron. Soc. 91, 483–490 (1931).





%\bibitem{Zelmanov_53} Зельманов А. Л.  Космология//Большая Советская Энциклопедия (Второе Издание). 1953 Т. 23. С. 109.






\end{thebibliography}
\end{document}